\documentclass{article}
\usepackage{amsmath}
\usepackage{amssymb}
\usepackage{epsfig}
\advance \oddsidemargin by -1cm
\advance \textwidth by -1cm
\advance \marginparwidth by 1cm

\newcommand{\ket}[1]{| #1 \rangle}
\newcommand{\bra}[1]{\langle #1 |}
\newcommand{\bracket}[2]{\langle #1 | #2 \rangle}
\newcommand{\qt}{\Gamma \backslash G}
\newcommand{\cj}{\overline}
\newcommand{\ci}{C^\infty(M)^*}
\newcommand{\tr}{\operatorname{tr}}
\begin{document}

\title{Spectral decompositions for evolution operators of mixing dynamical 
systems}
\author{S. Roberts and B. Muzykantskii \\ University of Warwick, \\ Coventry, 
UK, CV4 7AL}
\date{\today}
\maketitle

\begin{abstract}
  Spectral decompositions for the evolution operator on 
  an energy shell in  phase space are constructed for the free motion on 
compact 
2D surfaces of constant negative curvature. Applications to quantum chaos
  and in particular to the recently proposed ballistic $\sigma$-model are
  briefly discussed.
\end{abstract}

\section{Introduction}

  The dynamics of a Hamiltonian system can either be 
described in terms of the trajectories $x_0 \to x(t) = U_t x_0$ in the phase 
space, 
or 
by specifying the laws of evolution of a function $\varphi(x)$ on 
the
phase space: $ \varphi \to \hat U_t \varphi$. The evolution 
operator 
$\hat U_t $ which advances a function along the trajectories is defined by
\[
\hat U_t \varphi (x) = \varphi(U_t x) = \varphi (x(t)). 
\]

Quantum mechanics has a natural relation to the trajectory based approach
through Feynman's Path Integral. In the semiclassical limit the path
integral can be approximated by the saddle point contributions (which
turn out to be the classical orbits) and leads to the Gutzwiller trace
formula \cite{Gutzwiller} for the Green's function of the quantum
mechanical Hamiltonian. This is a useful tool for studying such problems as  the
quantum energy level correlations but it requires knowledge of the long
periodic orbits in order to obtain the correlations at small energy
differences~\cite{Berry,BogomolKeating}. The situation is particularly
bad in chaotic systems, where the periodic orbits proliferate
exponentially with length. In practice some uncontrolled approximations about
correlations of actions for different periodic orbits are made to get the
analytical results.

In contrast, the flow based approach seems suitable for studying the
behavior of chaotic systems at long times when due to  the decay of
correlations the dynamics becomes trivial. To use this approach in quantum
chaos two problems however need to be overcome. First, it is unclear how
to relate the quantum mechanics to the evolution operator since there is no
analogue of the Feynman path integral.  Second,  one needs to be able to 
calculate various properties of
the evolution operator which are naturally formulated in terms of the spectral 
decomposition of $\hat U_t$ in decaying eigenmodes.

The first problem has recently been addressed in~\cite{MK,AAAS} where it
has been conjectured that the (suitably averaged) correlation functions of
the quantum energy levels and/or quantum eigenfunctions can be generated
from an effective action of nonlinear $\sigma$-model type involving the
Liouville operator $\hat L = \frac{d}{dt} \hat U_t$ (the so called
``ballistic'' $\sigma$-model). The inspiration for this approach comes
from the well developed theory of weakly disordered metals, where the
disorder averaged properties of an ensemble of macroscopically identical
systems are calculated using a similar $\sigma$-model  with a diffusion
operator instead of the Liouvillian one.  (See \cite{Efetov,Efetov_book}
for a detailed discussion of the diffusive case). 

Our paper is devoted to the second problem. In particular we 
compute a generalized spectral decomposition of the evolution operator 
\begin{equation}\label{decomposition0}
\hat U_t = \sum_\lambda e^{\lambda t} \ket{\lambda} \bra{\tilde \lambda}.
\end{equation}
for a class of ``model'' chaotic systems namely the free motion on 
two dimensional compact  surfaces with constant negative curvature.

For mixing systems the only square integrable eigenfunctions of $\hat U_t$ are
the constant functions (see Appendix \ref{Mix}). The eigenfunctions 
$\ket{\lambda}$ entering
the spectral decomposition belong to a larger space $C^\infty(M)^*$ of
distributions  and can be obtained from
the residues of the analytical continuation of matrix elements of
the resolvent of $\hat L$ (see section \ref{general}). The procedure was 
suggested by Ruelle \cite{Ruelle} and was
successfully used to study the dynamics of some chaotic maps (see
\cite{Antoniou,Suchanecki,Antoniou_tent,Qiao}). The
eigenvalues $\lambda$ entering the spectral decomposition are sometimes called
``Ruelle resonances''.  For our model dynamical system the resonances can be 
found using 
the Selberg trace formula to relate the classical and quantum zeta-functions 
\cite{Rugh}.  This approach however does not provide the eigenmodes which we 
compute using the representation theory of $SO(2,1)$. 

The rest of the paper is organized as follows.  We start by outlining the
Ruelle procedure for  flows concentrating on criteria for convergence
of the resulting decomposition (section \ref{general}). We then
introduce geodesic flows on constant negative curvature surfaces (section
\ref{CNCS}); summarize the important facts from the representation
theory of $SO(2,1)$ and proceed to obtaining the spectral decomposition
(section \ref{special}, equation (\ref{SpecU})). This decomposition is
used to refine approximations of the decay of correlations for geodesic flows 
(section
\ref{decay}) and relate the evolution of particle density on the
configuration space to the Laplacian operator (section \ref{laplace}).
To conclude we discuss the  regularization of the Liouvillian operator entering 
the 
ballistic $\sigma$-model (section \ref{ssigma}).  Some technical details are 
relegated to the appendices.

\section{Decompositions of evolution operators for general mixing 
systems}\label{general}

\subsection{Spectral decompositions}

Consider a Hamiltonian system on a phase space $T^*N$ with coordinates $x=(q,p)$ 
where $N$ is a smooth compact manifold parameterised by q and p is the momentum.
When the Hamiltonian $H$ does not depend on
time  energy is conserved and the trajectories lie on surfaces $M$ of
constant energy. 
  We study the restriction of the flow $U_t$ to a constant energy shell $M$ 
for some value of the energy.  If $dH$ is non-zero
on $M$ the Liouvillian measure $dpdq$ induces a measure $d \mu$ on 
$M$ according to $d \mu d H =
dp dq$ \cite{Rugh}. By Liouville's theorem this measure is
preserved by the flow $U_t$.

The evolution operator $\hat U_t$ advances a function $\varphi(x): M \to
\mathbb C$
along the flow
\[
\hat U_t \varphi(x) = \varphi( U_t x)= e^{\hat L t} \varphi(x)
\]
where for a Hamiltonian system the Liouville operator $\hat L$ is given by
\begin{equation}\label{Liouville}
\hat L \varphi = \left. \frac{d}{dt} \right|_{t=0} \hat U_t \varphi =\{ H , 
\varphi \} 
\end{equation}
with $\{ \quad , \quad \}$ being the Poisson bracket.  

The operator $\hat U_t$ is unitary with respect to
the scalar product
\begin{equation}\label{prod}
\bracket{\xi(x)}{\varphi(x)} = \int_{x \in M} \xi(x) \cj{\varphi(x)} \,
d\mu(x)
\end{equation}
since $\mu$ is preserved by $ U_t$. A natural space for $\hat U_t$ is
the Hilbert space $L_2(M)$ of  square integrable functions on $M$,
while $\hat L$ preserves a smaller space of compactly supported
infinitely differentiable functions $C^\infty(M)$.
 
We would like to find a spectral decomposition (\ref{decomposition0}) for the 
operator $\hat U_t$ in
terms of its eigenvalues $e^{\lambda t}$ and projectors $\bra{\tilde \lambda}$ 
onto its eigenfunctions $\ket{\lambda}$.
This would enable us to establish the evolution of a function $\varphi(x)$:
\[
\hat U_t \ket{\varphi} = \sum_\lambda e^{\lambda t} \ket{\lambda}
\bracket{\tilde \lambda}{\varphi} 
\]
Since the operator $\hat U_t$ is unitary on $L_2(M)$ it can only have
eigenvalues which lie on the unit circle.  On the other hand, in a mixing
chaotic system (definitions and properties of mixing systems are given in
appendix \ref{Mix}) all deviations from a constant value decay implying the 
existence of modes corresponding to eigenvalues with modulus
less than 1.  The resolution of this apparent paradox lies in observing that 
the 
$L_2(M)$
eigenfunctions of $\hat U_t$ do not necessarily form a basis in $L_2(M)$
or even in $C^\infty(M)$. In fact, for a mixing system the only square
integrable eigenfunctions of $\hat U_t$ are the constant functions which
have the eigenvalue 1 (appendix \ref{Mix}).  In order to find a spectral 
decomposition for
$\hat U_t$ we need to extend the Hilbert space $L_2(M)$ to a rigged
Hilbert space as described below. A similar approach was successfully
employed for various maps (see e.g. \cite{Antoniou,Suchanecki}).

\subsection{Rigged Hilbert spaces}

A Hilbert space $\mathbb{H}$ may be extended to the set of linear
functionals on a suitable dense subspace $\mathbb S$.  The resulting
space $\mathbb S^*$ is called the rigging of $\mathbb{H}$ over $\mathbb
S$.  We denote by $f[\varphi]$ the value of the linear functional $f \in
\mathbb S^*$  on the vector $\varphi \in \mathbb S$.

A vector $g \in \mathbb H$ may be naturally embedded in $\mathbb S^*$ as
$g[\varphi] = \bracket{\varphi}{g}$, so we get a sequence of spaces $
\mathbb S \subset \mathbb{H} \subset \mathbb S^*$.  We make the above
embedding explicit~\cite{Antoniou} and denote the functional $f \in
\mathbb S^*$ by $\ket{f}$; its value on a vector $\varphi \in \mathbb S$
being $f[\varphi] = \bracket{\varphi}{f}$. We also introduce the notation
$\bracket{f}{\varphi}$ for the antilinear functional $\cj{f[\varphi]}$.

We search for  eigenfunctionals of the evolution operator $\hat U_t$ in
the rigging $\ci$ of $L_2(M)$ over $C^\infty(M)$.  The evolution operator
is extended to $\ci$ by
\[
\bracket{\varphi}{\hat U_t f} = \bracket{\hat U_{-t} \varphi}{f} \text{
  where } \ket{f} \in \ci \text{ and } \varphi \in C^\infty(M).
\]
Since $\hat U_t$ does not preserve any scalar product in $\ci$ its
eigenvalues need not lie on the unit circle.

We shall construct decompositions for the correlation function
\[
\bra{\xi} \hat U_t \ket{\varphi}=\int_M d \mu \xi(x) \hat U_t \varphi(x)
\]
in a subset of eigenfunctionals  $\{ 
\ket{f_\lambda}\}$  
from $\ci$:
\begin{equation}
  \label{decomposition1}
\bra{\xi} \hat U_t \ket{\varphi} = \sum_\lambda e^{\lambda t}
\bracket{\xi}{f_\lambda}\bracket{f_{-\cj{\lambda}}}{\varphi} , \hat U_t 
\ket{f_\eta} = e^{\eta t} \ket{f_\eta}.
\end{equation}
In general (\ref{decomposition1}) only has asymptotic meaning (see section 
\ref{details} for details), 
and will only converge  when $t>0$ and $\xi$ and $\varphi$ belong to 
a subspace  
$\mathbb{T}$ of $C^\infty(M)$. For the free motion on compact surfaces of 
constant negative curvature we find such a subspace which is dense in the set of 
infinitely differentiable functions. 
 Note, that the eigenfunction
$\ket{f_{-\cj{\lambda}}}$ with the eigenvalue $e^{-\cj{\lambda} t}$
appears in~(\ref{decomposition1}) due to the unitarity condition $(\hat
U_t)^+=U_{-t}$ which leads to another expansion for the correlation
function:
\begin{equation}
\label{decomposition-}
\bra{\xi} \hat U_t \ket{\varphi} =   
\cj{\bra{\varphi}\hat U_{-t} \ket{\xi}}=
 \sum_\lambda e^{-\cj{\lambda} t}
\bracket{\xi}{f_{-\cj{\lambda}}}\bracket{f_\lambda}{\varphi}
\end{equation}
converging when $t<0$ and $\xi,\varphi \in \mathbb{T}$

Decompositions (\ref{decomposition1},\ref{decomposition-}) appear naturally in
  connection with the correlation function $\bra{\xi}(\hat L
  -z)^{-1}\ket{\phi}$ of the resolvent of $\hat L$ .

\subsection{Resolvent method for calculating the evolution operator 
decompositions}\label{details}

Choosing an integral representation for the resolvent of the Liouville 
operator converging when $\operatorname{Re}z >0$
\begin{equation}\label{Resolvent}
R_-(z) = -\int_0^\infty e^{-zT} \hat U_T  \, dT = (\hat L -z)^{-1}
 \end{equation}
and assuming decomposition~(\ref{decomposition1}) we obtain for the
correlation function of the resolvent
\begin{equation}
\label{decomposition2}
F_{\xi,\varphi}(z) \equiv \bra{\xi}R_-(z)\ket{\varphi} = \sum_\lambda
\frac{\bracket{\xi}{f_\lambda}\bracket{f_{-\cj{\lambda}}}{\varphi}}{\lambda-z}
\end{equation} 
where $\xi,\varphi \in \mathbb{T}$.

Conversely decomposition~(\ref{decomposition1}), can be constructed by
analytically continuing $F$ to the left half of the z-plane and analyzing
its singularities and residues. 
The position of the poles dictate the rate of 
decay of the correlation function of the  evolution operator~\cite{Ruelle}.  
The 
values of $\lambda$ in decomposition (\ref{decomposition1}) are called Ruelle 
resonances.

Each term in decomposition~(\ref{decomposition1}) is well defined for 
$\xi,\varphi \in C^\infty(M)$ when 
\begin{eqnarray}\label{condn1}
&\quad& \text{the position of the poles of 
$F_{\xi,\varphi}(z)$ } \nonumber \\ &\quad& \text{does not  depend on the choice 
of $\xi$  or $\varphi$. }
\end{eqnarray}
A famous conjecture due to Ruelle \cite{Ruelle} states that the poles of the 
resolvent for a mixing system do indeed satisfy this condition.

The sum in (\ref{decomposition1}) converges if in addition the residues 
$\operatorname{Res}(\lambda,F_{\xi,\varphi})$ grow slowly enough  when 
$|\lambda| \to \infty$:
\begin{equation}\label{condn2}
\lim_{R \rightarrow \infty} \sum_{R<|\lambda|<\infty} e^{\lambda t}
\operatorname{Res}(\lambda,F_{\xi,\varphi}) =0   
\end{equation}
 The set $\mathbb{T}$ will comprise of functions in $C^\infty(M)$ which satisfy 
condition (\ref{condn2}).

To prove the convergence we consider the integral of $F_{\xi,\varphi,t}(z) = 
e^{zt}\bra{\xi} 
R_-(z)\ket{\phi}$ for $\xi$ and $\phi$ in $\mathbb{T}$ around the 
contour $|z|= R$ for a given large $R$.  In the limit $R
\rightarrow \infty$ this integral converges due to (\ref{condn2})
to
\begin{equation}\label{infctr}
\lim_{R \rightarrow \infty} \int_{|z|=R} F_{\xi,\varphi,t}(z) \, dz = 2\pi i
\sum_\lambda\operatorname{Res}(\lambda, F_{\xi,\varphi,t}) = 2\pi i
\sum_\lambda  e^{\lambda t} \operatorname{Res}(\lambda, F_{\xi,\varphi})
\end{equation}
where the sum is over all the poles $\lambda$ other than $\infty$.

Since all the poles are in the left half of the z-plane the contour can
be deformed to go along the line $\operatorname{Re}z = a$ for a fixed
$a>0$ leading to
\begin{eqnarray}
\int_{\operatorname{Re} z = a} F_t(z) \, dz &=& \int_{\operatorname{Re}z = a} 
e^{zt} \int_{T=0}^\infty e^{-zT} 
\bra{\xi} \hat U_T \ket{\varphi} \, dT \, dz \nonumber \\
 &=&  \int_{T=0}^\infty 
\int_{y=-\infty}^\infty e^{(a+iy)(t-T)} \,dy \bra{\xi} 
\hat U_T \ket{\varphi} \, dT \nonumber \\
&=& \int_{T=0}^\infty \delta(t-T) \bra{\xi} \hat U_T \ket{\varphi} \, dT 
\nonumber \\ &=&  \bra{\xi} \hat U_t \ket{\varphi}  \text{ for 
} 
 t > 0  
\end{eqnarray}

From equation~(\ref{infctr}) we get the 
decomposition 
into residues 
\begin{equation}\label{ExpRes}
\bra{\xi} \hat U_t \ket{\varphi} =  2\pi i \sum_\lambda e^{\lambda t} 
\operatorname{Res}(\lambda, F)
\end{equation}
which converges absolutely for $t > 0$.
 
The individual terms in (\ref{ExpRes}) exist for arbitrary $\xi,\varphi \in 
C^\infty(M)$ but the series converges only for $\xi,\varphi \in \mathbb{T}$.
The residue $\operatorname{Res}(\lambda,F)$ is a linear functional of $\xi$ 
and 
an antilinear functional of $\varphi$.  We define the operator 
$\hat K_\lambda: C^\infty(M) \to C^\infty(M)^*$ by
\[
2 \pi i \operatorname{Res}(\lambda, F) = \bra{\xi} \hat K_\lambda 
\ket{\varphi}  \quad \xi, \varphi \in C^\infty(M)
\]
so that
\[
\bra{\xi} \hat U_t \ket{\varphi} =  \sum_\lambda e^{\lambda t} \bra{\xi} \hat 
K_\lambda \ket{\varphi} \quad \xi, \varphi \in \mathbb{T}
\]
and $\hat K_\lambda\ket{\varphi}$ is an eigenfunctional of $\hat U_t$
\[
\hat U_t \hat K_\lambda\ket{\varphi} = e^{\lambda t} \hat 
K_\lambda\ket{\varphi}.
\]
Let $\{ \ket{f_\lambda^k} \}$ be a basis for the eigenspace corresponding
to the eigenvalue
$e^{\lambda t}$ i.e.  the image 
of $C^\infty(M)$  under $\hat K_\lambda$. 
\[
\hat K_\lambda \ket{\varphi} = \sum_k \bracket{c_k}{\varphi} 
\ket{f_\lambda^k} 
\text{ with } \hat U_t \ket{f_\lambda^k} = e^{\lambda t} \ket{f_\lambda^k}.
\]
Substituting this expression into (\ref{ExpRes}) we obtain 
\begin{eqnarray}\label{decomposition3}
\bra{\xi} \hat U_t \ket{\varphi} &=&  \sum_\lambda \bra{\xi} \hat K_\lambda 
\ket{\varphi} \nonumber \\
&=& \sum_\lambda \sum_k 
\bracket{\xi}{f_\lambda^k}\bracket{f_{-\cj{\lambda}}^k}{\varphi} \quad 
\xi,\varphi \in \mathbb{T} , t>0
\end{eqnarray}
where by virtue of the unitarity of the evolution operator the coefficients 
$\bracket{c_k}{\varphi}$ are given by eigenfunctionals  
$\ket{f_{-\cj{\lambda}}^k}$  of $\hat U_t$ with eigenvalues 
$e^{-\cj{\lambda} t}$.
\[
\bracket{c_k}{\varphi} = \cj{ \bracket{\varphi}{f^k_{-\cj \lambda}} }
\] 
Equation (\ref{decomposition3}) reduces to (\ref{decomposition1}) in the
non-degenerate case when the image of $\hat K_\lambda$ is one dimensional.

For arbitrary $\xi$ and $\phi$ in $C^\infty(M)$ the convergence of decomposition 
(\ref{decomposition1}) is asymptotic:
\begin{equation}\label{limpet}
\Big| \bra{\xi} U_t \ket{\varphi} -  \sum_{\operatorname{Re}\lambda \geq -a} 
e^{\lambda 
t} 
\bracket{f_{-\cj{\lambda}}}{\varphi} \bracket{\xi}{f_\lambda} \Big| < C(a) 
e^{-at} 
\quad ,  a>0.
\end{equation}
This inequality holds since the integral 
\[
\int_0^\infty \left( \bra{\xi} U_t \ket{\varphi} -  
\sum_{\operatorname{Re}\lambda \geq -a} e^{\lambda t} 
\bracket{f_{-\cj{\lambda}}}{\varphi} \bracket{\xi}{f_\lambda} \right ) e^{-zt} 
\, dt 
\]
is analytic in the region $\operatorname{Re} z > -a$  of the z-plane.

A similar procedure starting with the representation $R_+(z)$
for the resolvent converging at $\operatorname{Re} z<0$
\begin{equation}\label{Resolvent+}
R_+(z)= \int_{-\infty}^0 e^{zt}\hat U_t \, dt = (\hat L - z)^{-1} 
\end{equation}
gives the decomposition (\ref{decomposition-})  which converges absolutely
for $t<0$ when the conditions (\ref{condn1},\ref{condn2}) are met.

\subsection{Evolution operator at long times}\label{longt}

We can use the decompositions (\ref{decomposition1}) and 
(\ref{decomposition-})  
to study the approach to equilibrium at long times for mixing dynamical systems.

We show below that for $\xi$ and $\varphi$ in $\mathbb{T}$ the function $C(a)$ 
in (\ref{limpet}) is independent of $a$. 
The behavior at long future times may be approximated by retaining only
the terms in (\ref{decomposition1}) where $\lambda$ has 
a small negative real part.
\begin{eqnarray}
\bra{\xi} \hat U_t \ket{\varphi} &=& \sum_{\operatorname{Re}\lambda \geq -a} 
e^{\lambda t} 
\bracket{f_{-\cj{\lambda}}}{\varphi} \bracket{\xi}{f_\lambda}
+  \sum_{\operatorname{Re}\lambda <-a} e^{\lambda t} 
\bracket{f_{-\cj{\lambda}}}{\varphi} \bracket{\xi}{f_\lambda}
\nonumber \\
&\approx& \sum_{\operatorname{Re}\lambda \geq -a} e^{\lambda t} 
\bracket{f_{-\cj{\lambda}}}{\varphi} \bracket{\xi}{f_\lambda} \nonumber
\\
&=&  \bracket{\varphi}{1} \bracket{1}{\xi}  
+ \sum_{0 > \operatorname{Re}\lambda \geq -a} e^{\lambda t} 
\bracket{f_{-\cj{\lambda}}}{\varphi} \bracket{\xi}{f_\lambda} , \nonumber
\\
\end{eqnarray}
where we have  separated out the the contibution from the constant $L_2(M)$ 
eigenfunction with $\lambda =0$.  For a mixing
system the latter is the only non-decaying eigenfunction.
By the absolute convergence of the sum (\ref{decomposition1}) for $t>0$ the 
discarded 
terms are bounded 
by
\begin{eqnarray}
|\sum_{\operatorname{Re}\lambda <-a} e^{\lambda t} 
\bracket{f_{-\cj{\lambda}}}{\varphi} 
\bracket{\xi}{f_\lambda}| &\leq& 
e^{-at}\sum_{\operatorname{Re}\lambda <-a} | 
\bracket{f_{-\cj{\lambda}}}{\varphi} 
\bracket{\xi}{f_\lambda} |  < e^{-at} C \nonumber
\end{eqnarray}
for some constant C and hence decay faster than $e^{-at}$.

For long past times we work analogously from (\ref{decomposition-}) rather than 
(\ref{decomposition1}) to obtain:
\[
\bra{\xi} \hat U_t \ket{\varphi} \approx  \bracket{\varphi}{1} 
\bracket{1}{\xi} 
+ \sum_{0 < \operatorname{Re} (-\cj{\lambda}) \leq b} e^{-\cj{\lambda} t} 
\bracket{f_\lambda}{\varphi} \bracket{\xi}{f_{-\cj{\lambda}}} \text{ for
  } t \to -\infty.
\]
Here the terms which decay faster than $e^{bt}$ as $t \rightarrow -\infty$ 
have been discarded.

\section{Surfaces of constant negative curvature}\label{CNCS}

We will use the resolvent method described in the previous section to obtain a 
spectral decomposition for the 
evolution operator of  a 'model' chaotic system. The simplest 
systems which are strongly chaotic (and in particular mixing) are the free 
motion on compact 2D surfaces of constant negative curvature.

The hyperbolic plane $N$  plays the role of a 
universal cover for these surfaces.  It can be 
embedded in 
Minkowski space where the metric is  
\[
ds^2 = -dx_1^2+dx_2^2+dx_3^2
\]
as the surface satisfying the equation
\[
-x_1^2+x_2^2+x_3^2 = -1.
\]

We shall consider the free motion on compactifications of $N$
formed  by quotienting it under the action of some discrete group as described
below.  We start by analyzing the free motion of a particle on the hyperbolic 
plane itself. 
This dynamical system has the phase space $T^*N$ and is described by the 
Hamiltonian
$
H= \frac{p^2}{2m}
$
which is just the kinetic energy of the particle.
The trajectories for this dynamical system are given by the geodesics of the 
surface $N$.
For a particle with unit mass $m=1$, the constant energy surface
$M$ with energy $E=\frac{1}{2}$ consists of the points with momenta of unit 
modulus
(and hence also unit speed).  Thus the energy shell $M$ is the 
unit 
cotangent bundle of the hyperbolic plane.
Changing the energy amounts to a rescaling which leaves the trajectories 
unchanged and only alters the rate at which they are traversed.  

The group of isometries of the hyperbolic plane $G=SO(2,1)$ 
acts simply transitively on the points of the energy shell.
Let us take as a base point in $M$ the point $O=(q,p)$ 
where the position $q$ is given by $(0,0,1)$ and the momentum $p$ is 
$(1,0,0)$.
A point $x$ on the energy shell may be identified with the unique element of 
$G$ which takes  $O$ to $x=gO$.  In this way the constant energy shell
can be identified with the elements of $G$.  The above construction gives a 
diffeomorphism from the topological group G to the energy shell M.

We choose a basis for the Lie algebra $\boldsymbol{g}$ of G consisting of the 
matrices 
\begin{equation}\label{Lie}
P =\begin{pmatrix} 0 & 1 & 0 \\
                        1 &0 &0 \\
                        0 &0 & 0 
\end{pmatrix}    
\quad Q= \begin{pmatrix} 0 & 0 & 1 \\
         0 &0 &0 \\
         1 &0 & 0  \end{pmatrix}
         \quad K= \begin{pmatrix} 0 & 0 & 0 \\
                        0 &0 &1 \\
                        0 & -1& 0  \end{pmatrix}
\end{equation}
An element $g \in G$ may be written as a product \cite{Vilenkin} 
\begin{equation}\label{Euler}
g(\phi,\tau,\psi) = e^{K\phi} e^{P\tau} e^{K\psi} .
\end{equation}
The parameters $\tau$, $\phi$ and $\psi$ are known as the Euler angles.
Note that $\tau$ and $\phi$ give the position in polar coordinates: 
\[
(x_1,x_2,x_3) = (\cosh \tau, \sinh \tau \cos \phi, \sinh \tau 
\sin \phi)
\]
while the  angle $\psi$ gives a consistent way of parameterizing the 
direction of the momentum. 

After a time $t$ a free particle at the point $O$ on the constant energy 
surface advances along the geodesic to which it belongs to the point $h_tO$  
where $h_t = e^{Pt}$.
The point $gO$ moves to $gh_tO$; hence on identifying $M$ with $G$ the
evolution corresponds to right multiplication by the element 
$h_t$.
This 1-parameter group of transformations 
\[
g \mapsto U_t(g) = gh_t
\]
is the geodesic flow on G.  
The measure $\mu$ on $M$ which is invariant under the geodesic flow $U_t$
is the Haar measure $dg$ of the group G.  This measure is invariant under 
right 
and 
left multiplication by elements of the group.  The $U_t$ invariant scalar 
product 
for functions on G is therefore given by
\begin{eqnarray}
\bracket{f_1}{f_2} &=& \int_{g \in G} f_1(g) \cj{f_2(g)}  \, dg \nonumber \\
&=& \frac{1}{4\pi^2}\int_0^\infty\int_0^{2\pi}\int_0^{2\pi} 
f_2(\phi,\tau,\psi) \cj{f_2(\phi,\tau,\psi)}  \sinh \tau \,  d\tau \, 
d\phi \, d\psi. \nonumber
\end{eqnarray}

Finally we notice that right multiplication by elements in the rotation
subgroup $H = \{e^{K\psi}\}$ only affects the direction of momentum and
does not change the position of the particle.  Therefore we can identify
the hyperbolic plane $N$ with the set of right cosets $G/H$.

We now turn to the free motion on general surfaces of constant negative
curvature which are constructed by  taking a 
tessellation of the hyperbolic plane $G/H$ and identifying the tessellating 
shapes 
(the fundamental domains).
Let $\Gamma$ be the group of transformations mapping the tessellating shapes 
to 
each other.  For every $\gamma \in \Gamma$ the points
$\gamma gH$ and $gH$ of the hyperbolic plane are identified.  
The points of the quotient surface formed under this identification are 
labeled by double cosets in $\qt/H$. When the 
directions of momentum at each of the points are included the 
constant energy 
surface $M = \qt$ is formed. 
Since the free motion is given by the right shift by 
$h_t$ it is not  
affected by quotienting on the left by the subgroup $\Gamma$, so
on the constant energy surface 
$\qt$ the geodesic flow is given by 
\begin{equation}\label{gflow}
U_t(\Gamma g) = \Gamma gh_t
\end{equation}

\section{Decomposition of the evolution operator for the geodesic 
flows}\label{special}

In this section we use the representation theory of $G=SO(2,1)$ to find a
decomposition of the form (\ref{decomposition2}) for the resolvent for
the geodesic flow on $\qt$.

The right regular representation $T_R(h)$ of $G$ on $L_2(\qt)$ is defined by
\[
T_R(h) \varphi(\Gamma g) = \varphi(\Gamma gh) \text{ where } \varphi \in 
L_2(\qt).
\]

It can be decomposed as a direct sum of irreducible unitary representations 
$T^y$ which leads to a splitting of  $L_2(\qt)$ into a  direct sum of the 
spaces 
$\mathbb{H}(T^y)$ on 
which $T^y$ acts.
\begin{equation}\label{reduced}
L_2(\qt) = \underset{y \in Y}{\oplus} \Bbb{H}(T^y) 
\end{equation}

As the geodesic flow $U_t$ defined in (\ref{gflow}) amounts to right
multiplication by the group element $h_t$ the evolution operator $\hat
U_t$ coincides with $T_R(h_t)$ and leaves $\mathbb H(T^y)$ invariant
leading to the decomposition
\begin{equation}\label{EvolDec}
\hat U_t = \sum_{y \in Y} T^y(h_t) .
\end{equation}

Substituting (\ref{EvolDec}) into  (\ref{Resolvent}) we obtain the resolvent 
for the geodesic flow:
\begin{equation}\label{ResolDec}
R_-(z) = -\int_0^\infty e^{-zt}\hat U_t \, dt 
= \sum_{y \in Y} R_-^y(z)   
\end{equation}
where  $R_-^y(z) = -\int_0^\infty e^{-zt} T^y(h_t) \, dt$.
 
The rest of this section is organized as follows.
In section \ref{Irrep} we  discuss the unitary irreducible
representations $T^y$.  In section \ref{SpecL} we study the decompositions 
(\ref{reduced}) and relate it to the spectra of the Laplacian on the quotient 
surface $\qt/H$.  In section \ref{SpecDec} we calculate the
integrals $R_-^y(z)$ for each of these irreducible representations and  
in section \ref{SpecEvol} we  combine all the results to find $R_-(z)$.

\subsection{Irreducible representations of $SO(2,1)$}\label{Irrep}

Let T be an arbitrary unitary representation of $G=SO(2,1)$ on the Hilbert 
space 
$\Bbb{H}(T)$ 
with scalar product $\bracket{\quad}{\quad}$.
The Casimir operator is defined on $\Bbb{H}(T)$ by 
\[
\Omega(T) = L^2_P(T) + L^2_Q(T) -L^2_K(T)
\]
where $L_X(T)$ is the Lie derivative of $T$ in the direction $X \in
\boldsymbol{g}$ and $P, Q, K \in \boldsymbol g$ are as defined in (\ref{Lie}). 
This operator commutes with each of the $T(g)$~\cite{Ratner}
and therefore must be a scalar multiple of the identity on each of the 
irreducible representations $T^y$.
It is convenient to denote by $T^\rho$ the unitary irreducible representation 
on
which this scalar is $-\frac{1}{4}-\rho^2$, i.e. for all $ v \in 
\Bbb{H}(T^\rho)$ we have:
\begin{equation}\label{Casimir}
\Omega(T^\rho) v = (-\frac{1}{4}-\rho^2)  v
\end{equation}
The
following values of $\rho$ are allowed \cite{Ratner}:
\begin{itemize}
\item
$\operatorname{Im}\rho = 0$ and $\operatorname{Re}\rho \geq 0$ (the
principal series)   
\item
$\operatorname{Re}\rho = 0$ and $\operatorname{Im}\rho \in
(0,\frac{1}{2})$ (the complimentary series)
\item
$\operatorname{Re}\rho = 0$ and $\operatorname{Im}\rho \in \mathbb{N}$
each corresponding to a pair of inequivalent representations (the discrete 
series).
\end{itemize}
In addition we have the 1-dimensional identity representation $I$ for which 
$\Omega(I)v = 0$ for $v \in \mathbb H(I)$.

Under the action of the compact abelian subgroup $H = \{ e^{Kt} \}$ the 
representation $T^\rho$ splits into one dimensional irreducible 
representations. 
Let $\ket{n} \in \Bbb{H}(T^\rho)$ be a vector in one such one-dimensional 
representation with
\begin{equation}\label{n}
T^\rho(e^{Kt}) \ket{n} = e^{int} \ket{n}
\end{equation}
It is a fact \cite{Lang} that in a representation $T^\rho$ there is at most 
one vector $\ket{n}$ for each value $n$ and they can be normalized to form an
orthonormal basis for the space $\mathbb H(T^\rho)$.  We also use the notation 
$ 
\ket{\rho, n}$ when the irreducible representation to which $\ket{n}$ belongs 
needs to be 
specified.
For the representations of the principal and complimentary series the value of 
n ranges over all the integers. For the discrete series, one of the pair has a 
basis consisting of $\ket{n}$ where $n \geq \operatorname{Im} \rho$  and the 
other where $n \leq \operatorname{Im} \rho$~\cite{Lang}. 

A vector $\ket{\varphi}$ in $H(T^\rho)$ may be expanded in the 
basis $\{ \ket{n} \}$ as
\begin{equation}\label{nExp}
\ket{\varphi} = \sum_n \bracket{n}{\varphi} \ket{n}
\end{equation}
In the basis $\{ \ket{n} \}$ the matrix elements for the representation 
$T^\rho$ 
are  given by \cite{Vilenkin} 
\[
\bra{m} T^\rho(g) \ket{n} = 
e^{im\phi}e^{in\psi}B^l_{mn}(\cosh \tau) 
\]
where $l=\-\frac{1}{2}+i\rho$ and $B^l_{mn}(\cosh \tau)$ are the Jacobi 
functions.

\subsection{Decomposition of $L_2(\qt)$}\label{SpecL}

We now consider the 
decomposition of  $L_2(\qt)$ into irreducible representations $T^\rho$ for a 
particular quotient surface $\qt/H$.

Suppose that the decomposition (\ref{reduced}) of $L_2(\qt)$ has $N_\rho$ 
copies 
of the irreducible representation $T^\rho$, labeled by $s \in \{ 1, ... 
,N_\rho 
\}$:
\begin{equation}\label{red}
L_2(\qt) = \oplus_\rho (\oplus^{N_\rho}_{s=1} \mathbb H(T^{\rho;s})).
\end{equation}  
Each $\mathbb H(T^{\rho;s})$ has a basis $\{ \ket{\rho,n;s} \}$ as described 
in 
section (\ref{Irrep}).  From (\ref{Casimir}) and (\ref{n}) we deduce that the 
space spanned by the vectors $\{ \ket{\rho,n;s} \text{ where } s = 1, ... , 
N_\rho \}$ is the 
intersection of the eigenspaces of $\Omega(T_R)$ and $L_K(T_R)$ with the 
eigenvalues $-\frac{1}{4}-\rho^2$ and $in$ respectively.
\begin{eqnarray}\label{diff}
\Omega(T_R) \ket{\rho,n;s} &=& (-\frac{1}{4}-\rho^2) \ket{\rho,n;s} \nonumber 
\\
L_k(T_R) \ket{\rho,n;s} &=& in \ket{\rho,n;s} 
\end{eqnarray}

The functions $\ket{\rho,0;s}$ are invariant under multiplication on the right 
by elements of $H$ and therefore can be viewed as functions in $L_2(\qt/H)$.  
On 
this space the Casimir operator $\Omega$ reduces to the Laplacian $\hat 
\Delta$ 
hence 
the values of $\rho$ occuring in (\ref{red}) correspond to the eigenvalues 
$\epsilon(\rho) = -\frac{1}{4}-\rho^2$ of the Laplacian $\hat \Delta$ on the 
quotient 
surface $\qt/H$. $N_\rho$ is the multiplicity of the eigenvalue 
$\epsilon(\rho)$ and $\ket{\rho,0;s}$ are its eigenfunctions.
It is known that the spectrum $\Lambda_\Gamma$ of $\hat \Delta$ on a compact 
surface $\qt/H$ is discrete and is bounded above by 0.  It also contains the 
eigenvalue 0 which corresponds to the the constant eigenfunction. Hence  
(\ref{red}) takes the form:
\begin{equation}\label{red2}
L_2(\qt) = \underset{-\frac{1}{4}-\rho^2 \in \Lambda_\Gamma}{\oplus} 
(\oplus_{s=1}^{N_\rho} \Bbb{H}(T^{\rho 
;s})).
\end{equation}

A differential equation for the eigenfunctions $\ket{\rho,n;s}$ is obtained by 
unwrapping a function in $L_2(\qt)$ to give a periodic function on the whole 
group $\chi: G \to 
\mathbb C$ which obeys
\[
\chi(\gamma g) = \chi(g) \text{ for all } \gamma \in \Gamma.
\] 
Using Euler's coordinates (\ref{Euler}) on $G$ in (\ref{diff}) we deduce that 
the unwrapping  of the function $\ket{\rho,n;s}$ has the form 
$\chi_s^{\rho,n}(\tau,\theta) e^{in\psi}$ where $\chi_s^{\rho,n}(\tau,\theta)$ 
obeys the second order differential equation: 
\begin{equation}\label{diff2}
 \left(  \frac{1}{\sinh 
\tau}\frac{\partial}{\partial\tau} \sinh\tau 
\frac{\partial}{\partial\tau} + \frac{1}{\sinh^2 \tau} \left(
\frac{\partial^2}{\partial^2\phi} -2in \cosh \tau \frac{\partial}{\partial 
\phi} 
-n^2 \right) \right) \chi_s^{\rho,n} 
= (-\frac{1}{4}-\rho^2) \chi_s^{\rho,n} . 
\end{equation}

\subsection{Spectral decompositions for the irreducible 
representations}\label{SpecDec}

We now find the correlation functions of the  resolvents $R_-^\rho(z)$
for each of the irreducible 
representations $T^\rho$ and obtain a spectral decomposition for the 
operator $T^\rho(h_t)$.

Expanding
\[
\bra{\xi} R_-^\rho(z) \ket{\varphi} = -\int_0^\infty e^{-zt} \bra{\xi} 
T^{\rho}(h_t) \ket{\varphi} \, dt
\]
in the basis $\{\ket{n}\}$ we get
\begin{equation}\label{Resol}
\bra{\xi} R_-^\rho(z) \ket{\varphi} =
 \int_0^\infty e^{-zt} \sum_{m,n}  \bracket{\xi}{m} \bra{m} 
T^{\rho}(h_t) \ket{n} \bracket{n}{\varphi} \, dt  
\end{equation}
The matrix element $\bra{m} T^{\rho}(h_t) \ket{n}$ can be written as a sum 
over the exponentials $e^{\lambda t}$ where 
\begin{equation}\label{lambda}
\lambda \in  
\{l, l-1, l-2 ...; \cj l, \cj l-1,\cj l-2,...  \} = 
\Lambda_\rho
\end{equation}
and
\begin{equation}\label{matelts}
\bra{m} T^{\rho}(h_t) \ket{n} = B^l_{mn}(\cosh t) = 
\sum_{\lambda \in \Lambda_\rho}  c_{mn}^{\lambda} e^{\lambda t}
\end{equation}
Substituting (\ref{matelts}) into the resolvent (\ref{Resol}) we obtain
\begin{eqnarray}\label{Resol2}
\bra{\xi} R_-^\rho(z) \ket{\varphi} &=& \sum_{\lambda \in \Lambda_\rho}
 \int_0^\infty e^{(\lambda-z)t} \sum_{m,n}  \bracket{\xi}{m} c^\lambda_{mn}
 \bracket{n}{\varphi} \, dt \nonumber \\
 &=& \sum_{\lambda \in \Lambda_\rho} \sum_{m,n} \frac{ \bracket{\xi}{m} 
c^\lambda_{mn} \bracket{n}{\varphi} }{\lambda - z} \nonumber \\
 &=& \sum_{\lambda \in \Lambda_\rho}  \frac{ \bra{\xi} \hat 
K_\lambda \ket{\varphi} }{\lambda - z}. 
 \end{eqnarray}
The operators $\hat K_\lambda$ determining  the residue at the pole $\lambda 
\in 
\Lambda_\rho$ are 
given by
\[
\hat K_\lambda = \sum_{m,n} \ket{m} c^\lambda_{mn} \bra{n}.
\]
The matrix elements  $c_{mn}^\lambda$ may be split into a product
(appendix \ref{Coeffs}) $c_{mn}^{\lambda} = a_m^{\lambda} \cj{b_n^{\lambda}}$
which enables us to write (\ref{Resol2}) in the form
\begin{eqnarray}\label{resolq}
\bra{\xi} R_-^\rho(z) \ket{\varphi} 
&=& \sum_{\lambda \in \Lambda_\rho} \frac{\underset{n}{\sum} \bra{n} 
\cj{b_n^\lambda} \ket{\varphi} \underset{m}{\sum} \bra{\xi} a^\lambda_m 
\ket{m}}{\lambda - z} \nonumber \\
&=& \sum_{\lambda \in \Lambda_\rho} 
\frac{\bracket{f_{-\cj{\lambda}}}{\varphi}\bracket{\xi}{f_\lambda}}{\lambda - 
z}
\end{eqnarray}
where $\lambda$ runs through the set $\Lambda_\rho$ (\ref{lambda}) and
\[
\ket{f_\lambda} = \sum_m a^{\lambda}_m \ket{m} \text{  ,  }
\ket{f_{-\cj{\lambda}}} = \sum_m b^{\lambda}_m \ket{m} 
\]
are eigenvectors of $T^\rho(h_t)$:
\[
T^\rho(h_t) \ket{f_\eta} = e^{\eta t} \ket{f_\eta} .
\]
The coefficients $a_m^\lambda$ and $b_m^\lambda$ where $\lambda = l, \cj l$ 
are 
found explicitly in appendix \ref{Coeffs} along with a procedure for 
generating 
the eigenfunctionals for the other values of $\lambda \in \Lambda_\rho$.
 
The eigenfunctionals  $\ket{f_\lambda}$ and $\ket{f_{-\cj{\lambda}}}$
are linear functionals on the space $\mathbb{S}^\rho \subset \mathbb{H}(T^\rho)$
 of test vectors $\ket{\varphi} = \sum_n c_n \ket{\rho, n} \in \mathbb 
H(T^\rho)$ 
where the coefficients $c_n \to 0$ as $|n| \to \infty$ faster than any power of 
n i.e. 
$\lim_{|n| \to \infty} c_n n^k = 0 \text{ for all } k \in \mathbb{N}$.
By comparing (\ref{resolq}) with (\ref{decomposition2}) we arrive at the 
spectral decomposition for $T^\rho(h_t)$
\begin{equation}\label{SpecT}
T^\rho(h_t) = \sum_{\lambda \in \Lambda_\rho} e^{\lambda t} 
\ket{f_\lambda} 
\bra{f_{-\cj{\lambda}}}.
\end{equation}
This decomposition converges absolutely for $t>0$ for test vectors in the dense 
subspace $\mathbb{T}^\rho \subset \mathbb{S}^\rho$ of vectors of the form
$\ket{\varphi} = \sum_{n=-K}^{K} c_n \ket{\rho, n} \in \mathbb{H}(T^\rho)$ for 
some 
$K$.
(appendix \ref{Conv}).

\subsection{Spectral decompositions of the evolution operators for the 
geodesic 
flows}\label{SpecEvol}

On combining (\ref{SpecT}) with (\ref{red2}) 
we obtain the central result of this paper: a spectral decomposition 
for $\hat U_t$:
\begin{equation}\label{SpecU}
\hat U_t  = T_R(h_t) = \sum_{-\frac{1}{4}-\rho^2 \in \Lambda_\Gamma} 
\sum_{\lambda \in \Lambda_\rho} \sum_{s=1}^{N_\rho}  
 e^{\lambda t} 
\ket{f_\lambda^{\rho;s}} 
\bra{f_{-\cj{\lambda}}^{\rho;s}} 
\end{equation}
where $\Lambda_\Gamma$ is the spectrum of the Laplacian on $\qt/H$ containing 
eigenvalues $-\frac{1}{4}-\rho^2$ with multiplicities $N_\rho$ and where
$\Lambda_\rho$  given by equation (\ref{lambda}) and also shown in figure 
\ref{poles}.  The 
functionals 
\begin{eqnarray}
\ket{f_\lambda^{\rho;s}}  &=& \sum_n a^{\lambda}_n 
\ket{\rho, n; s} \nonumber \\
\ket{f_{-\cj{\lambda}}^{\rho;s}} &=& \sum_n b^{\lambda}_n 
\ket{\rho, n; s} \nonumber
\end{eqnarray}
are the eigenfunctionals of $\hat U_t$:
\[
\hat U_t \ket{f_\eta^{\rho;s}} = e^{\eta t} \ket{f_\eta^{\rho;s}}.
\]
The functions $\ket{\rho,n;s}$, $s=1, ... N_\rho$ form an orthogonal basis in 
the space of solutions of  the linear differential equation (\ref{diff2}).  
The 
coefficients are obtainable by expanding $B^l_{mn}(\cosh t)$ in powers of 
$e^{\lambda t}$ (see equation \ref{matelts}) and are given explicitly for 
$\lambda 
= l, \cj l$ in (\ref{sputnik}).

The eigenfunctionals $\ket{f_\eta^{\rho;s}}$ belong to the space 
$C^\infty(\qt)^*$ and act on test functions in $\underset{-\frac{1}{4}-\rho^2 
\in \Lambda_\Gamma}{\oplus} \mathbb{S}^\rho = C^\infty(\qt)$.
The decomposition (\ref{SpecU}) converges for functions in a dense subspace   
\begin{equation}\label{subT}
\mathbb T = \{ \ket{\varphi}  =  \sum_{-\frac{1}{4}-\rho^2 \in \Lambda_\Gamma} 
 \sum_{s=1}^{N_\rho} \sum_{n=-K}^{K} c_{n,\rho;s} \ket{\rho,n;s} \text{ for some 
} K \}  \subset C^\infty(\qt)
 \end{equation}
 of test functions whose unwrappings have only a finite number of Fourier 
components in the angle $\psi$
(appendix \ref{Conv}).

The eigenvalues entering (\ref{SpecU}) could also be obtained by comparing 
classical orbit expansions for the traces of the resolvents of $\hat L$ and 
$\hat \Delta$ \cite{Rugh_evals}.

\section{Properties of the evolution operators for the geodesic flows}

\subsection{Decay of correlations}\label{decay}

At long times only the leading terms with $k=0$ in each of the $\Lambda_\rho$  
in the 
spectral 
decomposition for $\hat U_t$ (\ref{SpecU}) remain significant (see section 
\ref{longt}).
Therefore as $t \rightarrow \infty$, for $\xi,\varphi \in \mathbb{T} \subset 
C^\infty(\qt)$ 
\[
\bra{\xi} \hat U_t \ket{\varphi}  \approx \sum_{-\frac{1}{4}-\rho^2 \in 
\Lambda_\Gamma} 
 \sum_s  e^{(-\frac{1}{2} \pm i\rho) t} 
\bracket{f^{\rho;s}_{\frac{1}{2}\pm i\rho}}{\varphi} 
\bracket{\xi}{f_{-\frac{1}{2}\pm i\rho}^{\rho;s}} 
\]
Separating out the eigenvalue with $\lambda=0$  for a compact quotient surface 
where the 
dynamics is mixing we find that
\begin{equation}\label{ldec}
\hat U_t \ket{\varphi}  \approx \bracket{1}{\varphi} \bracket{\xi}{1} +  \sum_{0 
\neq 
-\frac{1}{4}-\rho^2 \in \Lambda_\Gamma} 
\sum_s  e^{(-\frac{1}{2} \pm i\rho) t} 
\bracket{f_{\frac{1}{2}\pm i\rho}^{\rho;s}}{\varphi} 
\bracket{\xi}{f_{-\frac{1}{2}\pm i\rho}^{\rho;s}} 
\end{equation}

\begin{figure}
\begin{center}
\epsfig{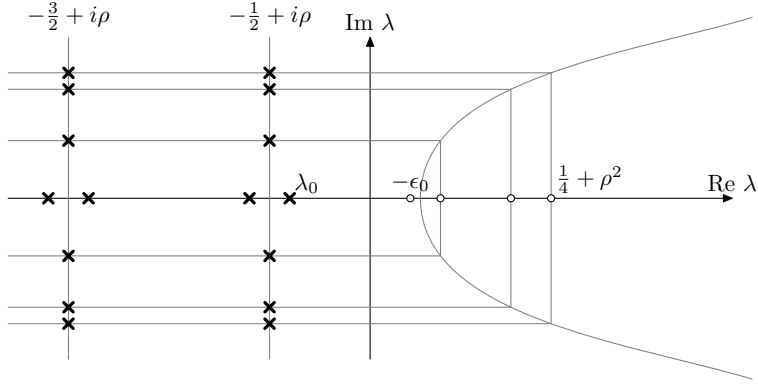}
\caption{The poles $\lambda = -\frac{1}{2}-i\rho-k$ (crosses) of the resolvent 
  give the eigenvalues $e^{\lambda t}$ of the evolution operator
  appearing in decomposition (\ref{SpecU}).  These poles are related to
  the eigenvalues $-\epsilon = \frac{1}{4}+\rho^2$ of $-\hat \Delta$
  (open circles). The rate of decay $e^{\lambda_0 t}$ at long times is
  determined by the pole $\lambda_0$ with the least negative real part.}
\label{poles}
\end{center}
\end{figure}

The approach to equilibrium is governed by the eigenvalue(s) $e^{(-\frac{1}{2} 
\pm i\rho) t}$ in this expansion
 with the largest modulus (see figure \ref{poles}). 
There are two possibile cases distinguished by the smallest non-zero 
eigenvalue $-\epsilon_0$ of  $-\hat \Delta$.
\begin{enumerate}
\item
If $-\epsilon_0 < \frac{1}{4}$  then the slowest decaying term gives the decay 
rate $e^{\lambda_0 t}$ where $\lambda_0 =  
-\frac{1}{2}+\sqrt{\frac{1}{4}+\epsilon_0}$.
 For the non-constant part of the 
correlation function $\bra{\xi} \hat U_t \ket{\varphi}$ between  $\xi , 
\varphi 
\in C^\infty(\qt)$ we have
\begin{equation}\label{decay1}
\bra{\xi}\hat U_t \ket{\varphi} - \bracket{\xi}{1}\bracket{1}{\varphi} = 
O(e^{\lambda_0 t})
\end{equation}
\item
If $-\epsilon_0 > \frac{1}{4}$ then all the values of $\rho$ in (\ref{ldec}) 
are 
real and we get $\lambda_0 = -\frac{1}{2}$
\begin{equation}\label{decay2}
\bra{\xi}\hat U_t \ket{\varphi} - \bracket{\xi}{1}\bracket{1}{\varphi} = 
O(e^{-\frac{t}{2}})
\end{equation}
\end{enumerate}

By applying the result (\ref{limpet}) to the geodesic flows we find that for 
arbitrary $\xi, \varphi \in C^\infty(M)$
\[
|\bra{\xi} \hat U_t \ket{\varphi} - \bracket{\xi}{1} \bracket{1}{\varphi}| < 
Ce^{\lambda_0 t}
\]
This is a classical result proving the 
exponential decay of correlations for geodesic flows on constant negative 
curvature surfaces  (see 
e.g.~\cite{Ratner} for a detailed discussion).
Also from (\ref{limpet}) we get the other terms in the long time asymptote:
\begin{eqnarray}\label{asymp}
\Big| \bra{\xi} \hat U_t \ket{\varphi} &-&  \bracket{\xi}{1}\bracket{1}{\varphi}  
\nonumber \\ &-& \sum_{-\frac{1}{4}-\rho^2 \in 
\Lambda_\Gamma} 
 \sum_{s=1}^{N_\rho} \sum_{\pm} \sum_{k=0}^N 
 e^{(-\frac{1}{2}\pm i\rho-k) t} 
\bracket{\xi}{f_{-\frac{1}{2}\pm i\rho-k}^{\rho;s}} 
\bracket{f_{\frac{1}{2}\pm i\rho+k}^{\rho;s}}{\varphi} \Big| \nonumber \\ &<& 
C(N) 
e^{(\lambda_0 -(N+1))t} .
\end{eqnarray}

\subsection{Relation to the Laplacian}\label{laplace} 

In addition to refining the long time asymptotic form of the correlation 
function
(\ref{asymp}) the full expansion (\ref{SpecU}) provides a more 
detailed 
description of the evolution of distribution functions.
Consider a smooth function $\Psi$ on $\qt$ which is independent of the 
momentum 
coordinate $\psi$ (hence belongs to the space $\mathbb{T}$).  As it evolves 
under $\hat U_t$ this function will acquire 
some angular dependence but we will only be interested in the density on the 
configuration space i.e. the projection $\Psi_t = \hat P \hat U_t \Psi$ where 
$\hat P$ is the projector $\qt \to \qt/H$: 
\[
\hat P \chi = \frac{1}{2\pi} \int_0^{2\pi} \chi \, d\psi
\]  

Expanding $\Psi$ as a linear combination of the eigenfunctions $\ket{\rho 
,0 ;s}$ of the Laplacian for the quotient surface 
\[
\ket{\Psi} = \sum_{-\frac{1}{4}-\rho^2 \in \Lambda_\Gamma} \sum_s 
\bracket{\rho, 0, s}{\Psi} \ket{\rho ,0 ;s} 
\]
and using
\begin{eqnarray}
\hat U_t \ket{\rho,0;s} &=& T^{\rho}(h_t) \ket{\rho,0;s} 
\nonumber \\
&=& \sum_n \ket{\rho, n;s} \bra{\rho, n;s} T^{\rho}(h_t) 
\ket{\rho ,0;s} \nonumber \\
&=& \sum_n  B^{-\frac{1}{2}+i\rho}_{n0}(\cosh t) \ket{\rho, n;s} 
\nonumber
\end{eqnarray}
combined with the following expressions for the matrix elements of the 
projection 
operator 
\begin{eqnarray}
\hat P \ket{\rho, n;s} &=& 0 \quad  n \neq 0 \nonumber \\
\hat P \ket{\rho, n;s} &=& \ket{\rho, 0;s}  \nonumber
\end{eqnarray} 
we obtain for $\Psi_t$:
\begin{equation}\label{projection}
\ket{\Psi_t} = \hat P \hat U_t \ket{\Psi} =  \sum_{-\frac{1}{4}-\rho^2 \in 
\Lambda_\Gamma} B^{-\frac{1}{2}+i\rho}_{00}(\cosh t)
\sum_s \bracket{\rho, 0, s}{\Psi}  
 \ket{\rho ,0 ;s} 
\end{equation}

Using the symbolic notation
\[
\sqrt{ -\hat \Delta -\frac{1}{4}} \ket{\rho ,0 ;s} = \rho 
\ket{\rho ,0 ;s}
\] 
we rewrite (\ref{projection}) as
\begin{equation}\label{relop}
\hat P \hat U_t \ket{\Psi} =  B^{-\frac{1}{2}+i\sqrt{- \hat \Delta 
-\frac{1}{4}}}_{00}(\cosh t) \ket{\Psi}
\end{equation}
which relates the Laplacian $\hat \Delta$   on 
the 
surfaces 
of constant negative curvature to the classical dynamics in the configuration 
space.

\section{Conclusions}

\subsection{Summary of results}

Decompositions for the evolution operators $\hat U_t$ of a general mixing 
dynamical system on a manifold $M$ were constructed following Ruelle's 
prescription \cite{Ruelle} using the analytic 
continuation of the resolvent  (\ref{Resolvent}):
\[
F_{\xi,\varphi}(z) = -\int_0^\infty e^{-zT} \bra{\xi} \hat U_T \ket{\varphi} \, 
dT.
\]

It was proved that when conditions (\ref{condn1}, \ref{condn2}) are met we 
have the 
following decompostions of $\hat U_t$:
\begin{eqnarray}
\bra{\xi} \hat U_t \ket{\varphi} &=& \sum_\lambda e^{\lambda t} \sum_k 
\bracket{f^k_{-\cj \lambda}}{\varphi}\bracket{\xi}{f^k_\lambda} \text{ 
 convergent for $t>0$ } \nonumber \\
\bra{\xi} \hat U_t \ket{\varphi} &=& \sum_\lambda e^{-\cj \lambda t} \sum_k 
\bracket{f^k_\lambda}{\varphi}\bracket{\xi}{f^k_{-\cj \lambda}} \text{ 
 convergent for $t<0$ }
\nonumber
\end{eqnarray}  
where  $\ket{f^k_\eta}$ are eigenfunctionals of $\hat U_t$
\[
\hat U_t \ket{f^k_\eta} = e^{\eta t} \ket{f^k_\eta}
\]
belonging to the space $C^\infty(M)^*$. 
They can be obtained from the residues of $F_{\xi,\varphi}(z)$ which are 
positioned at $\lambda$ and have the form
\[
2\pi i \operatorname{Res}(\lambda,F_{\xi,\varphi}) =  \sum_k 
\bracket{\xi}{f^k_\lambda}\bracket{f^k_{-\cj \lambda}}{\varphi}.
\]

On applying this method to the evolution operator $\hat U_t$ for the free 
motion 
on a compact surface of 
constant negative curvature $\qt/H$ we obtained the decomposition 
(\ref{SpecU}) which is convergent for $\xi$ and $\varphi$ in a dense subspace 
$\mathbb{T} \subset C^\infty(\qt)$ of functions whose unwrappings onto functions 
of $G$ have only a finite number of Fourier harmonics in the Euler angle $\psi$ 
(see equations \ref{Euler}, \ref{subT}).
From this decomposition we obtained a refinement for the rate of decay of 
correlations (\ref{asymp}) for these systems. 

We also found that the projection of the evolution operator $\hat P \hat U_t$ 
from $\qt$ to $\qt/H$ 
is related to the Laplacian $\hat \Delta$ on the surface $\qt/H$ by:
\[
\hat P \hat U_t = B_{00}^{-\frac{1}{2}+i\sqrt{-\hat \Delta-\frac{1}{4}}}(\cosh 
t).
\] 

\subsection{Consequences for the ballistic $\sigma$-model}\label{ssigma}

Finally we discuss some consequences of the spectral decomposition
(\ref{decomposition1}) for the ballistic $\sigma$-model \cite{MK,AAAS}. 
Without
diving into (the still controversial) issue of the $\sigma$-model
derivation and its region of validity we quote below the result for the
effective action (see \cite{MK,AAAS,AAAS2} for a detailed discussion):
\begin{equation}
\label{eq:sigma-action}
S= \int d\mu \tr \Lambda W^{-1} \left\{  \hat L + i (\omega +i0)
  \Lambda \right\} W  ,
\end{equation}
where $W(x)$ belongs to some (super)-group and $\Lambda$ is a particular
matrix in this group obeying $\Lambda^2=1$. The detailed structure of the
target space is somewhat involved and for the purpose of our discussion
it  suffices to consider a toy model with $W \in SU(2)$ and
$\Lambda=\mbox{diag}(1,-1)$. Then the target space is a two-dimensional
sphere $S^2 = SU(2)/U(1)$ parametrized by the matrices $Q=W^{-1} \Lambda
W$. The action~(\ref{eq:sigma-action}) does not depend on the
parametrization of $S^2$ in terms of $W$. Indeed, introducing the
two-form $Q_* V$ on the energy shell $M$ obtained by pulling back the
invariant volume $V$ on $S^2$ by the function $Q(x): M \to S^2$ we get
for the action:
$$
S= \int_M (Q_* V) \wedge p dq + i(\omega + i0) \tr (\Lambda Q) d \mu,
$$
where $p dq$ is the antiderivative of the simplectic structure $dp
\wedge dq$.  If $Q_* V$ is exact (and it is certainly closed) the first
term does not depend on the choice of the antiderivative but only on the
simplectic structure $dp \wedge dq$.

It is universally believed that some sort of regularization should
supplement the effective action~(\ref{eq:sigma-action}).  It was
conjectured \cite{AAAS} that in the limit of a vanishing regulator the
eigenvalues of the regularized Liouvillian operator approach the Ruelle
resonances (which were refered to as ``eigenvalues of the Perron
Frobenius operator'' in~\cite{AAAS}). 

We would like to point out, that this conjecture must be further
clarified due to the existence of the two inequivalent sets of Ruelle
resonances.  One set of resonances is in the left half of the complex
plane, while the other is in the right.  These two sets originate from
the different branches of the resolvent given by the two integral
representations (equations~\ref{Resolvent} and \ref{Resolvent+} respectively).
As a result there exists two non-equivalent regularizations of the
operator $\hat L$.  Denoting by $L_{reg}$ the regularization of $\hat L$ with 
eigenvalues close to the Ruelle resonances in the left half plane and
observing that the operator $-L_{reg}^+$ has the eigenvalues close to the
Ruelle resonances in the right half plane we suggest that the two
regularizations of $L$ are $L_{reg}$ and $-L_{reg}^+$. 

We suggest the following structure for the regularized operator in the target
space:
\begin{equation}
\label{regularisation}
\hat L  + i(\omega+i0) \Lambda \longrightarrow  
\left(
  \begin{array}{cc} L_{reg} + i\omega  & 0 \\ 0 & -L_{reg}^+ - i\omega
  \end{array} \right).
\end{equation}
which ensures the convergence of the action (\ref{eq:sigma-action}).
Expanding the action~(\ref{eq:sigma-action}) near $Q=\Lambda$ as $W=1+
\left(
  \begin{array}{cc}0 & w \\ - \bar w & 0
  \end{array} \right)$ 
we verify that the choice (\ref{regularisation}) ensures that the quadratic
part of the action $\delta ^2 S$  is non-postively
defined:
$$
\delta^2 S  = - \int d \mu (\bar w,  w) \left( \begin{array}{cc} 0 & L_{reg} 
\\
    -L_{reg}^+ & 0 \end{array} \right) \left( \begin{array}{c} w \\ \bar w
  \end{array} \right)
$$
Note, that a similar structure for the regularized action  appeared in
the model with diffusive scattering \cite{BMM2}.

\section*{Acknowledgements} \nonumber

One of the authors (B. Muzykantskii) wishes to acknowledge useful discussions 
with D. Khmelnitskii and M. Pollicott.

\appendix
\section{Mixing dynamical systems}\label{Mix}
A dynamical system $(M, U_t, \mu)$ is mixing if for any two sets $A, B \subset 
M$
\begin{equation}\label{mixdef}
\lim_{t \to \infty} \mu(U_t A \cap B) = \mu(A) \mu(B)
\end{equation}

For a mixing system the only square integrable eigenfunctions of $\hat U_t$ 
are 
the constant functions which have the eigenvalue 1.  The proof \cite{Arnold} 
of 
this 
result is reproduced below.

We write (\ref{mixdef}) in terms of the characteristic functions  $\chi_A$ and 
$\chi_B$ of the sets A and B (The characteristic function $\chi_C$ of a set C 
takes on 
the value $\chi_C(x) =1$ if $x \in C$ and  0 otherwise).
\begin{equation}\label{mix1}
\lim_{t \to \infty} \bracket{\hat U_t \chi_A}{\chi_B} = 
\bracket{\chi_A}{1}\bracket{1}{\chi_B}
\end{equation}
Since the space of linear combinations of characteristic functions is dense in 
the space of square integrable 
functions (\ref{mix1}) must also be true for $f,g \in L_2(M)$
\begin{equation}\label{mix2}
\lim_{t \to \infty} \bracket{\hat U_t f}{g} = \bracket{f}{1}\bracket{1}{g}
\end{equation}
Let $f$ be an eigenfunction of $\hat U_t$ with eigenvalue $e^{\lambda t}$ and 
$g=1$ then (\ref{mix2}) becomes
\[
e^{\lambda t} \bracket{f}{1} = \bracket{f}{1}
\] 
therefore the eigenvalue must be 1.

\section{Coefficients for the eigenfunctionals}\label{Coeffs}

We now study the matrix elements $c^{\lambda}_{mn} = \bra{m} \hat K_\lambda 
\ket{n}$ of the operator $\hat K_\lambda$ appearing in the residues of the 
resolvent (\ref{Resol2})
\[
\bra{\xi}R_-^\rho(z)\ket{\varphi} = \sum_{\lambda \in \Lambda_\rho} 
\frac{ \bra{\xi} \hat K_\lambda \ket{\varphi} }{\lambda - z} =\sum_{\lambda 
\in 
\Lambda_\rho} \sum_{m,n}
\frac{ \bracket{\xi}{m} c_{mn}^\lambda \bracket{n}{\varphi} }{\lambda - z}.
\]

Consider the following operators on $\mathbb{H}(T^\rho)$ 
\begin{eqnarray}
L &=& L_P(T^\rho) \nonumber \\
B_- &=& L_K(T^\rho) + L_Q(T^\rho) \nonumber \\
B_+ &=& L_K(T^\rho) - L_Q(T^\rho). \nonumber 
\end{eqnarray}
Note that $T^\rho(h_t) = e^{Lt}$ and we have the commutation relations 
\begin{eqnarray}\label{commutation}
[L , B_+] &=& B_+ \nonumber \\ \relax
[L , B_-] &=& -B_- \nonumber \\ \relax
[B_-, B_+] &=& 2L \relax
\end{eqnarray}

From the commutation relations we have that $B_+ L = (L+1) B_+$ therefore $B_+ 
e^{Lt} = e^{(L+1)t} B_+$ i.e. $B_+ T^\rho(h_t) = e^t T^\rho(h_t) B_+$.
By considering  $\bra{\xi} B_+ R_-^\rho \ket{\varphi}$ and using this result 
we 
obtain
\begin{equation}\label{bplus}
B_+ \hat K_\lambda = \hat K_{\lambda+1} B_+
\end{equation}
Hence  $B_+$ sends the image of $\hat K_{\lambda}$ into the image of $\hat 
K_{\lambda+1}$.
In particular $B_+ \hat K_l =0 = B_+ \hat K_{\cj l} =0$ since there are no 
eigenvalues $l+1$ and $\cj l + 1$ in the set $\Lambda_\rho$.

Consider an eigenfunctional $\ket{f}$ in the image of $K_\lambda$ so that it 
satisfies  $T^\rho(h_t) \ket{f} = e^{\lambda t} \ket{f}$ and therefore
\[
L \ket{f} = \lambda \ket{f} .
\]
Suppose also that it is annihilated by $B_+$:
\[
B_+ \ket{f} = 0 
\]
so that using (\ref{commutation}) we get
\begin{eqnarray}
\bra{n} B_+ \ket{f} &=& 0 \nonumber \\
\bra{n} B_+ B_- \ket{f} &=& -2\bra{n} L \ket{f}  = -2\lambda \bracket{n}{f} 
\nonumber
\end{eqnarray}
for all $n \in \mathbb{Z}$.

Representing $\ket{f}$ by the linear combination 
\[
\ket{f} = \sum_n \alpha_n \ket{n} \text{ where } \alpha_n = \bracket{n}{f} 
\]
and using the explicit expression for the matrix elements of $B_+$ and $B_-$ 
\cite{Vilenkin}  
\begin{eqnarray}\label{Bpm}
-2i\bra{n} B_+ &=& -(l + n)\bra{n-1}  - 2n\bra{n} 
           +(l-n)\bra{n+1} \nonumber \\
-2i\bra{n} B_- &=& (l+n)\bra{n-1}- 2n\bra{n} 
           -(l-n)\bra{n+1} 
\end{eqnarray}
 we get 
the 
system of linear equations 
\begin{equation}
\begin{split}
&-(l + n)\alpha_{n-1}  - 2n \alpha_{n} +(l-n)\alpha_{n+1} = 0 \nonumber \\
&-(l+n-1)(l+n)\alpha_{n-2} -2(l+n) \alpha_{n-1} +2( l(l+1) +n^2)\alpha_n 
\nonumber \\
&- 2(l-n)\alpha_{n+1} -(l-n-1)(l-n)\alpha_{n-2} = -4\lambda \alpha_n \nonumber
\end{split}
\end{equation}

These equations have only the trivial solution $\alpha_n =0$  
unless $\lambda = l = -\frac{1}{2} + i\rho$ or $\lambda = \cj l = -\frac{1}{2} 
- 
i\rho$  in 
which case they have the unique 
non-trivial solutions 
\begin{eqnarray}\label{lead}
\alpha_n^l &=& \frac{1}{(l+n)!(l-n)!} \nonumber \\
\alpha_n^{\cj l} &=& (-1)^n
\end{eqnarray} 
respectively.
Hence the images of $\hat K_l$ and $\hat K_{\cj l}$  are one-dimensional.

Since $B_+$ sends the image of $\hat K_\lambda$ into the image 
of $\hat K_{\lambda+1}$  and its restriction on the image of 
$\hat K_\lambda$ has zero kernel for $\lambda  \neq l,\cj l$ we deduce that 
the images of  $\hat K_{l-k}$ and $\hat K_{\cj l-k}$ for all $k \in 
\mathbb{N}$  
are also one-dimensional.
The operators $\hat K_\lambda$ where $\lambda \in \Lambda_\rho$ may therefore 
be 
written as 
\[
\hat K_\lambda = \ket{f_\lambda} \bra{f_{-\cj{\lambda}}}  = \sum_{m,n} \ket{m} 
a_m^\lambda \cj{b_n^\lambda} \bra{n} = \sum_{m,n} \ket{m} c_{mn}^\lambda 
\bra{n}
\]
and their matrix elements $c_{mn}^\lambda$ split into a product 
$c_{mn}^\lambda= 
a_m^\lambda \cj{b_n^\lambda}$ where
\begin{eqnarray}\label{spud}
\ket{f_\lambda} &=& \sum_m a_m^\lambda \ket{m} \nonumber \\
\ket{f_{-\cj{\lambda}}} &=& \sum_n b_n^\lambda \ket{n}
\end{eqnarray}
as stated in the main text.

Combining (\ref{lead}) with (\ref{spud}) we see that $a_n^l = \alpha_n^l$ and 
$a_n^{\cj l} = \alpha_n^{\cj l}$.  
This can be confirmed from equation 
(\ref{matelts}) and the expansion of $B^l_{m0}(\cosh t)$ at 
large t \cite{Magnus}:
\begin{eqnarray}\label{Jacobi}
B^l_{m0}(\cosh t) &=& \Big( \frac{2^l (l!)^2}{\sqrt{\pi}} 
\frac{1}{(l+m)!(l-m)!} e^{lt} \nonumber \\
&+& \frac{2^{\cj l} (l!) }{\sqrt{\pi}} (-1)^m  
e^{\cj l t}\Big) \nonumber \\
&\times& \Big(1 + O(e^{-t})\Big).
\end{eqnarray}
Using (\ref{Jacobi}) and the symmetry relation \cite{Vilenkin}
\[
B^l_{mn}(\cosh t) = \frac{(l+n)!(l-n)!}{(l+m)!(l-m)!} B^l_{nm}(\cosh t)
\] 
we can find the coefficients $b_n^l$ and $b_n^{\cj l}$:
\begin{eqnarray}
b_n^l &=& \frac{2^l}{\sqrt{\pi}} (l!)^2 \nonumber \\
b_n^{\cj l} &=& \frac{2^{\cj l}}{\sqrt{\pi}} (l!)^2 (l+n)!(l-n)! \nonumber
\end{eqnarray}

To summarize the functionals $\ket{f_\lambda} , \ket{f_{-\cj \lambda}}$ 
involved 
in the decomposition of $\hat K_\lambda$ 
for $\lambda= l, \cj l$ have the following form
\begin{eqnarray}\label{sputnik}
\ket{f_l} &=& \sum_n \frac{1}{(l+n)!(l-n)!} \ket{n} \nonumber 
\\
\ket{f_{\cj l}} &=& \sum_n (-1)^n \ket{n} \nonumber \\
\ket{f_{-\cj l}} &=& \sum_n \frac{2^l}{\sqrt{\pi}} (l!)^2 \ket{n} \text{ and } 
\nonumber 
\\
\ket{f_{-l}} &=& \sum_n \frac{2^{\cj l}}{\sqrt{\pi}} (l!)^2 (l+n)!(l-n)! 
\ket{n} 
\end{eqnarray}

Given the eigenfunctionals $\ket{f_l}$ and $\ket{f_{-\cj l}}$ the  
eigenfunctionals $\ket{f_\lambda}$ for the other values of $\lambda \in 
\Lambda_\rho$ may be 
obtained  by successive applications of the operator $B_-$:
\begin{equation}\label{ladder-}
\ket{f_{\lambda-k}} = B_-^k \ket{f_\lambda} 
\end{equation}
due to the relation $B_- \hat K_\lambda = \hat K_{\lambda -1} B_-$ which is 
proved 
analogously to (\ref{bplus}).

Similarly, the projectors can be obtained by applying the operator $B_+$ to 
$\ket{f_{-\cj l}}$ and $\ket{f_{-l}}$:
\begin{equation}\label{ladder+}
\ket{f_{-\cj\lambda+k}} = B_+^k \ket{f_{-\cj\lambda}} .
\end{equation}

From (\ref{Bpm}) and the relations (\ref{ladder-},\ref{ladder+}) we find that
\begin{eqnarray}\label{est}
a_n^{l-k} &=& \bracket{n}{f_{l-k}} = \bra{n} B_-^k \ket{f_l} = O(n^k) \nonumber 
\\ 
a_n^{\cj l-k} &=& \bracket{n}{f_{\cj l-k}} = \bra{n} B_-^k \ket{f_{\cj l}} = 
O(n^k) \nonumber \\ 
b_n^{l-k} &=& \bracket{n}{f_{-\cj l+k}} = \bra{n} B_+^k \ket{f_{-\cj l}} = 
O(n^k) \nonumber \\ 
b_n^{\cj l-k} &=& \bracket{n}{f_{-l+k}} = \bra{n} B_+^k \ket{f_{- l} }= O(n^k)  
\end{eqnarray}

\section{Convergence of the spectral decompositions}\label{Conv}

We now present the necessary condition for the convergence of
the spectral decomposition (\ref{SpecT}):
\begin{equation}\label{EvT}
\bra{\xi} T^\rho(h_t) \ket{\varphi} = \sum_{\lambda \in \Lambda_\rho} 
e^{\lambda 
t} \bracket{f_{-\cj{\lambda}}}{\varphi}\bracket{\xi}{f_\lambda}.
\end{equation}
In particular we show it is convergent when   $\xi$ and $\varphi$ belong to the 
dense subspace 
$\mathbb T^\rho$ of $ S^\rho \subset \mathbb H(T^\rho)$ where
\begin{equation}\label{Tsub}
\mathbb T^\rho = \{ \ket{\varphi} = \sum_{n=-K}^{K} c_n 
\ket{\rho, n} \text{ for some } K \}
\end{equation}

First we show that each term $e^{\lambda 
t} \bracket{f_{-\cj{\lambda}}}{\varphi}\bracket{\xi}{f_\lambda}$ in (\ref{EvT}) 
is defined for $\xi$ and $\varphi$ in the subspace $\mathbb S^\rho$ :
\begin{equation}\label{Ssub}
\mathbb{S}^\rho = \{ \ket{\varphi} = \sum_n c_n 
\ket{\rho,n} \text{ where } \lim_{|n| \to \infty} c_n n^q = 0 \text{ for all }  
q \in \mathbb{N} \}.
\end{equation}
Note that $\mathbb S^\rho$ is invariant under $T^\rho(h_t)$.
Let $\xi$ and $\varphi$ be given by 
\begin{eqnarray}\label{sums}
\ket{\varphi} &=& \sum_n c_n \ket{\rho,n} \nonumber \\ 
\ket{\xi} &=& \sum_n d_n \ket{\rho,n} 
\end{eqnarray}

Using 
$\bracket{f_{-\cj{\lambda}}}{m}\bracket{n}{f_\lambda} = a_m^\lambda 
\cj{b_n^\lambda}$ we get
\begin{equation}\label{term}
|e^{\lambda 
t} \bracket{f_{-\cj{\lambda}}}{\varphi}\bracket{\xi}{f_\lambda}| <  
 \sum_{m,n}|e^{\lambda t} a_m^\lambda \cj{b_n^\lambda} c_m \cj{d_n}|
\end{equation}
The eigenvalue $\lambda$ is of the form $l -k$ or $\cj l -k$ so by (\ref{est})
$a_m^\lambda = O(m^k)$ and $b_n^\lambda = O(n^k)$.
Hence
\[
|e^{\lambda 
t} \bracket{f_{-\cj{\lambda}}}{\varphi}\bracket{\xi}{f_\lambda}| <  
 C|e^{\lambda t}| \sum_m  |c_m m^k| \sum_n |\cj{d_n} n^k| 
\]
which converges due to (\ref{Ssub}).  Therefore each term is defined for 
$\xi,\varphi \in 
\mathbb{S}^\rho$ and the eigenfunctionals $\ket{f_\eta}$ belong to the space 
$\mathbb S^{\rho*}$.

Now we show that the sum (\ref{EvT}) converges for $\xi,\varphi \in \mathbb 
T^\rho$.  Since the sums (\ref{sums}) for $\xi,\varphi$ have only a finite 
number of terms it suffices to prove convergence for $\ket{\varphi} = \ket{n}$ 
and $\ket{\xi} = \ket{m}$. In this case (\ref{EvT}) reduces to (\ref{matelts}) 
which converges absolutely for $t>0$.

Now we turn to the convergence of the spectral decomposition of the evolution 
operator $\hat U_t$ (\ref{SpecU})
\begin{equation}\label{EvU}
\bra{\xi} \hat U_t \ket{\varphi} = \sum_{-\frac{1}{4}-\rho^2 \in 
\Lambda_\Gamma} 
 \sum_{s=1}^{N_\rho}  \sum_{\lambda \in \Lambda_\rho}
 e^{\lambda t} 
\bracket{\xi}{f_\lambda^{\rho;s}} 
\bracket{f_{-\cj{\lambda}}^{\rho;s}}{\varphi} 
\end{equation}
By the above considerations the terms in this series $e^{\lambda t} 
\bracket{\xi}{f_\lambda^{\rho;s}} 
\bracket{f_{-\cj{\lambda}}^{\rho;s}}{\varphi}$ are defined for $\xi,\varphi \in 
\underset{-\frac{1}{4}-\rho^2 \in \Lambda_\Gamma}{\oplus} \mathbb S^\rho = 
C^\infty(\qt)$.
  
We will show that  expansion (\ref{EvU}) is convergent when $\xi,\varphi$ 
belong to the  subspace $ \mathbb T \subset C^\infty(\qt)$:
\begin{equation}\label{sub}
\mathbb T = \{ \ket{\varphi}  =  \sum_{-\frac{1}{4}-\rho^2 \in \Lambda_\Gamma} 
 \sum_{s=1}^{N_\rho} \sum_{n=-K}^{K} c_{n,\rho;s} \ket{\rho,n;s} \text{ for some 
} K \}
\end{equation}
 
It is sufficient to determine the convergence for
\begin{eqnarray}
\ket{\varphi} &=&  \sum_{-\frac{1}{4}-\rho^2 \in \Lambda_\Gamma} 
 \sum_{s=1}^{N_\rho}  c_{\rho;s} \ket{\rho,n;s} \nonumber \\
\ket{\xi} &=&  \sum_{-\frac{1}{4}-\rho^2 \in \Lambda_\Gamma} 
 \sum_{s=1}^{N_\rho}  d_{\rho;s} \ket{\rho,m;s} \nonumber 
\end{eqnarray}
for which  decomposition (\ref{EvU}) is
\begin{equation}\label{spot}
\bra{\xi} \hat U_t \ket{\varphi} = \sum_{-\frac{1}{4}-\rho^2 \in 
\Lambda_\Gamma} 
\sum_{s=1}^{N_\rho}    \sum_{\lambda \in \Lambda_\rho}   
 e^{\lambda t} c_{\rho,s} a_n^{\lambda} \cj{d_{\rho,s} b_m^{\lambda}}   
\end{equation}

Let  $\varphi(\tau,\theta, \psi) = \varphi_n(\tau,\theta) e^{in\psi}$ be the 
unwrapping of $\ket{\varphi}$.
The function $\varphi_n$  on $\qt/H$ has the norm
\[
\|\varphi_n\|^2 = \frac{1}{2\pi} \int_F |\varphi_n(\tau,\theta)|^2 \sinh \tau 
\, 
d\tau \, d\theta \text{ , F is a fundamental domain }
\]

We may expand $\varphi_n$  in the eigenfunctions 
$\chi_n^{\rho;s}$ (the unwrappings of $\ket{\rho,n;s}$) 
of the operator (\ref{diff2}):
\begin{equation}\label{Exp2}
\varphi_n(\tau,\theta) =  \sum_{-\frac{1}{4}-\rho^2 \in \Lambda_\Gamma} 
 \sum_{s=1}^{N_\rho} c_{\rho,s} \chi_n^{\rho;s}.
\end{equation}
hence 
\[
\| \varphi_n \|^2 =  \sum_{-\frac{1}{4}-\rho^2 \in \Lambda_\Gamma} 
 \sum_{s=1}^{N_\rho} |c_{\rho;s}|^2 .
\]
Similarly
\[
\| \xi_n \|^2 =  \sum_{-\frac{1}{4}-\rho^2 \in \Lambda_\Gamma} 
 \sum_{s=1}^{N_\rho} |d_{\rho;s}|^2 . 
\]
Using the fact that
\[
\sum_{-\frac{1}{4}-\rho^2 \in 
\Lambda_\Gamma} 
\sum_{s=1}^{N_\rho} | c_{\rho,s} \cj{d_{\rho,s}} | < \|\xi_n\|\|\varphi_n \|
\]
and that (\ref{matelts}) converges absolutely  we see from (\ref{spot}) that
\begin{eqnarray}
|\bra{\xi} \hat U_t \ket{\varphi}| &=&  \sum_{-\frac{1}{4}-\rho^2 \in 
\Lambda_\Gamma} 
\sum_{s=1}^{N_\rho} | c_{\rho,s} \cj{d_{\rho,s}} | \sum_{\lambda \in 
\Lambda_\rho}   
  | a_n^{\lambda}\cj{b_m^{\lambda}} e^{\lambda t} | 
   \nonumber \\ 
&<&  
 C \sum_{-\frac{1}{4}-\rho^2 \in 
\Lambda_\Gamma} 
\sum_{s=1}^{N_\rho} | c_{\rho,s} \cj{d_{\rho,s}} | < C\|\xi_n\|\|\varphi_n \|  
\nonumber 
\end{eqnarray}
is bounded.

Hence taking $\xi,\varphi \in \mathbb{T}$ ensures the convergence   of the 
spectral decomposition (\ref{EvU}).  Since the unwrapping of $\ket{\rho,n;s}$ 
has the form $\chi_n^{\rho;s}$ the unwrappings of functions in $\mathbb T$ will 
have only a finite number of Fourier components in $\psi$.

\bibliographystyle{unsrt}
\bibliography{SRobeRef}

\begin{thebibliography}{10}

\bibitem{Gutzwiller}
M.~C. Gutzwiller.
\newblock Periodic orbits and classical quantization conditions.
\newblock {\em Journal of Mathematical Physics}, 12:343, 1971.

\bibitem{Berry}
M.~V. Berry.
\newblock Quantum chaology.
\newblock {\em Proceedings of the Royal Society of London A}, 413:183--198,
  1987.

\bibitem{BogomolKeating}
E.~B. Bogomolny and J.~P. Keating.
\newblock Gutzwiller's trace formula and spectral statistics: Beyond the
  diagonal approximation.
\newblock {\em Physical Review Letters}, 77(8):1472--1475, 1996.

\bibitem{MK}
Boris~A. Muzykantskii and David~E. Khmelnitskii.
\newblock Effective action in theory of quasi-ballistic disordered conductors.
\newblock {\em JETP letters}, 62(1):76--82, 1995.
\newblock Pis'ma ZH. Eksp Teor Fiz. 62 No 1, 68-74 (1995).

\bibitem{AAAS}
A.~V. Andreev, O.~Agam, B.~D. Simons, and B.~L. Altshuler.
\newblock Quantum chaos, irreversible classical mechanics and random matrix
  theory.
\newblock {\em Physical Review Letters}, 76:3947, 1996.

\bibitem{Efetov}
K.~B. Efetov.
\newblock Supersymmetry and theory of disordered metals.
\newblock {\em Advances in Physics}, 32(1):53--127, 1983.

\bibitem{Efetov_book}
K.~B. Efetov.
\newblock {\em Supersymmetry in Disorder and Chaos}.
\newblock New York: Cambridge University Press, 1997.

\bibitem{Ruelle}
Ruelle D.
\newblock Resonances of chaotic dynamical systems.
\newblock {\em Physical Review Letters}, 56(5):405--407, 1986.

\bibitem{Antoniou}
I.~Antoniou and S.~Tasaki.
\newblock Generalized spectral decompositions of mixing dynamical systems.
\newblock {\em International Journal of Quantum Chemistry}, 46:425--474, 1993.

\bibitem{Suchanecki}
S.~Tasaki Z.~Suchanecki, I.~Antoniou and O.F. Bandtlow.
\newblock Rigged hilbert spaces for chaotic dynamical systems.
\newblock {\em J. Math. Phys.}, 37(11):5837--5847, 1996.

\bibitem{Antoniou_tent}
I.~Antoniou and B.~Qiao.
\newblock Spectral decomposition of the tent maps and the isomorphism of
  dynamical systems.
\newblock {\em Physics Letters A}, 215:280--290, 1996.

\bibitem{Qiao}
B.~Qiao and I.~Antoniou.
\newblock Spectral decomposition of chebyshev maps.
\newblock {\em Physica A}, 233:449--457, 1996.

\bibitem{Rugh}
H.~H. Rugh.
\newblock Dynamical approach to temperature.
\newblock {\em Physical Review Letters}, 78(5):772--774, 1997.

\bibitem{Vilenkin}
N.J. Vilenkin.
\newblock {\em Special Functions and the Theory of Group Representations}.
\newblock Providence, RI: American Mathematical Society, 1968.

\bibitem{Ratner}
Marina Ratner.
\newblock The rate of mixing for geodesic and horocycle flows.
\newblock {\em Ergodic Theory Dynamical Systems}, 7(2):267--288, 1987.

\bibitem{Lang}
S.~Lang.
\newblock {\em $SL_2(\Bbb{R})$}.
\newblock Berlin: Springer, 1985.

\bibitem{Rugh_evals}
H.H. Rugh.
\newblock Generalized fredholm determinants and selberg zeta functions for
  axiom a dynamical systems.
\newblock {\em Ergodic Theory and Dynamical Systems}, 16(Pt4):805--819, 1996.

\bibitem{AAAS2}
A.~V. Andreev, B.~D. Simons, O.~Agam, and B.~L. Altshuler.
\newblock Semiclassical field theory approach to quantum chaos.
\newblock {\em Nuclear Physics B}, 482:536--566, 1996.

\bibitem{BMM2}
Y.M. Blanter, A.D. Mirlin, and B.A. Muzykantskii.
\newblock Quantum billiards with surface scattering: ballistic sigma-model
  approach.
\newblock {\em Physical Review Letters}, 80(19):4161--4164, 1998.

\bibitem{Arnold}
Arnold and Avez.
\newblock {\em Ergodic Problems of Classical Mechanics}.
\newblock New York: Benjamin, 1968.

\bibitem{Magnus}
W.~Magnus and F.~Oberhettinger.
\newblock {\em Formulas and theorems for the special functions of mathematical
  physics}.
\newblock Berlin: Springer, 1966.

\end{thebibliography}

\end{document}